\documentclass[11pt,a4paper]{article} 
\pdfoutput=1
\usepackage{mathtools}
\usepackage[utf8x]{inputenc}     % if unicode errors are coming
\usepackage{jheppub}
\usepackage{enumerate}
\usepackage{epsfig} 
\usepackage{float} 
\usepackage[caption = false]{subfig}
\usepackage{wasysym} 
\usepackage{mathrsfs} 
\usepackage{graphicx} 
\usepackage{amsfonts} 
\usepackage{amsbsy} 
\usepackage{amscd} 
\usepackage{multirow} 
\usepackage{tikz}
\usepackage{color}
\usepackage{slashed}
\usepackage{array}
\usepackage{tablefootnote}
\usepackage{booktabs}
\usepackage{caption}
\usepackage{subcaption}
\usepackage{adjustbox}
\usepackage{longtable}
\usepackage{orcidlink}  % ORCID ID

\usetikzlibrary{arrows,positioning,shapes.geometric} 
\usepackage[compat=1.1.0]{tikz-feynman}          % package allowing Feynman diagrams
\usepackage[font=small,labelfont=bf]{caption}
\usepackage{slashed}
\usepackage{multirow}
\usepackage{tabularx}
\usepackage{soul}  % Strikethrough text with \st{Hellow world}
\usepackage{listings} 
\usepackage{color}
\usepackage{amsthm}

\usepackage{amsthm}

\newcommand{\im}{\text{Im}}

\title{
Stable and Interpretable Jet Physics with IRC-Safe Equivariant Feature Extraction

}

\author[a]{Partha Konar\,\orcidlink{0000-0001-8796-1688}\,}
\author[b]{Vishal~S.~Ngairangbam\,\orcidlink{0000-0002-7143-715X}\,}
\author[b]{Michael~Spannowsky\,\orcidlink{0000-0002-8362-0576}\,}
\author[a, c]{and Deepanshu Srivastava\,\orcidlink{0009-0008-6796-0459}\,}
\affiliation[a]{Theoretical Physics Division, Physical Research Laboratory,\\ Shree Pannalal Patel Marg, Ahmedabad - 380009, Gujarat, India}
\affiliation[b]{Institute for Particle Physics Phenomenology, Department of Physics, Durham University,\\ Durham DH1 3LE, United Kingdom}
\affiliation[c]{Discipline of Physics, Indian Institute of Technology, Palaj,\\ Gandhinagar - 382424, Gujarat, India}
\emailAdd{konar@prl.res.in}
\emailAdd{vishal.s.ngairangbam@durham.ac.uk}
\emailAdd{michael.spannowsky@durham.ac.uk}
\emailAdd{deepanshu@prl.res.in}

\abstract{
Deep learning has achieved remarkable success in jet classification tasks, yet a key challenge remains: understanding what these models learn and how their features relate to known QCD observables. Improving interpretability is essential for building robust and trustworthy machine learning tools in collider physics. To address this challenge, we investigate graph neural networks for quark–gluon discrimination, systematically incorporating physics-motivated inductive biases. In particular, we design message-passing architectures that enforce infrared and collinear (IRC) safety, as well as E(2) and O(2) equivariance in the rapidity–azimuth plane. Using simulated jet datasets, we compare these networks against unconstrained baselines in terms of classification performance, robustness to soft emissions, and latent representation structures. Our analysis shows that physics-aware networks are more stable across training instances and distribute their latent variance across multiple interpretable directions. By regressing Energy Flow Polynomials onto the leading principal components, we establish a direct correspondence between learned representations and established IRC-safe jet observables. These results demonstrate that embedding symmetry and safety constraints not only improves robustness but also grounds network representations in known QCD structures, providing a principled approach toward interpretable deep learning in collider physics.
}

\preprint{IPPP/25/58}

\allowdisplaybreaks

\begin{document}
\maketitle
	
\flushbottom

\section{Introduction} 

Deep learning has significantly advanced many areas of science, including high-energy physics, by enabling the extraction of complex, task-relevant features directly from low-level data~\cite{Albertsson:2018maf,Radovic:2018dip, Guest:2018yhq, Larkoski:2017jix,Bourilkov:2019yoi,Karagiorgi:2021ngt,Butter:2022rso,Huerta:2022kgj,Belis:2023mqs,Allaire:2023fgp,ParticleDataGroup:2024cfk}. In jet physics, neural networks have achieved remarkable performance in discriminating between jets initiated by quarks and gluons, or in identifying boosted heavy particles that decay hadronically~\cite{Cogan:2014oua,Komiske:2016rsd, Kasieczka:2017nvn,Qu:2019gqs,Kasieczka:2019dbj, Louppe:2017ipp, Fraser:2018ieu,Dreyer:2020brq,Moreno:2019bmu,Mikuni:2020wpr,Mikuni:2021pou,pmlr-v162-qu22b,Romero:2023hrk,Sahu:2023uwb,Leigh:2024ked,Woodward:2024dxb,Birk:2024knn,Geuskens:2024tfo,Mikuni:2025tar}. Despite these advances, a central challenge remains: interpreting what these models learn and relating their internal representations to established physics concepts.

Interpretability is particularly important in high-energy physics~\cite{deOliveira:2015xxd,Chang:2017kvc,Kasieczka:2020nyd,Agarwal:2020fpt,Builtjes:2022usj,Chen:2022pzc,Bradshaw:2022qev,Li:2022xfc,Ngairangbam:2023cps,Araz:2023mda,Metodiev:2023izu,Vent:2025ddm,Bhattacharya:2025ave}, where physical insight and theoretical control are essential for robust scientific conclusions. Unlike hand-engineered observables, deep learning models are data-driven and can, in principle, exploit subtle correlations and detector effects that may not correspond to known physics. This raises questions about the trustworthiness, generalisation, and physical insight of the features extracted by neural networks.

An approach to enhance interpretability is to incorporate \emph{inductive biases} that encode known symmetries or constraints of the physical system into the network architecture~\cite{Butter:2017cot,Bogatskiy:2020tje,Haddadin:2021mmo,Barenboim:2021vzh,Bogatskiy:2022hub,Gong:2022lye,Sahu:2024sts,Brehmer:2024yqw,Spinner:2025prg}. For jet substructure, two such important biases are infrared and collinear (IRC) safety~\cite{Komiske:2018cqr,Dolan:2020qkr,Shimmin:2021pkm,Konar:2021zdg,Shen:2023ofd,Konar:2023ptv,Athanasakos:2023fhq,Chatterjee:2024pbp,Bhardwaj:2024djv} and symmetry under transformations in the rapidity-azimuth plane, specifically E(2) (Euclidean group) and O(2) (Orthogonal group) equivariance. Imposing these constraints ensures that the learned features respect fundamental properties of Quantum Chromodynamics (QCD) and detector geometry, helping bridge the gap between learned representations and analytic observables.

Graph neural networks (GNNs)~\cite{1555942,mpnn,Shlomi:2020gdn,Duarte:2020ngm,Esmail:2024jdg} have emerged as a robust framework for representing jets as sets or graphs of particles, naturally accommodating permutation invariance and allowing the incorporation of physics-motivated message passing schemes. By designing GNNs that enforce IRC safety and group equivariance, it becomes possible to constrain the space of learnable functions to those compatible with established jet substructure observables, such as Energy Flow Polynomials (EFPs)~\cite{Komiske:2017aww,Cal:2022fnm}.

In this work, we demonstrate that embedding strong physics-motivated inductive biases, in particular infrared and collinear (IRC) safety, as well as equivariance~\cite{Athanasakos:2023fhq,Bhardwaj:2024djv} under E(2) or O(2), into graph neural networks significantly enhances both their robustness and interpretability for quark–gluon discrimination~\cite{PhysRevLett.65.1321,Gallicchio:2010sw,Gallicchio:2011xq,FerreiradeLima:2016gcz,Gras:2017jty,Komiske:2018vkc,Kasieczka:2018lwf,Reichelt:2021svh,Bright-Thonney:2022xkx}. These inductive biases constrain the networks to respect fundamental QCD properties and underlying symmetries, ensuring that the learned features capture physically meaningful aspects of jet substructure. To assess their impact, we systematically compare architectures that incorporate these constraints with those that do not, highlighting how the inclusion of physics principles shapes the internal organisation of learned representations.

We conduct this investigation using simulated jet datasets, training multiple message passing GNNs and analysing their graph-level latent representations through principal component analysis (PCA). This analysis reveals a clear distinction between equivariant and non-equivariant architectures. The explained variance rises faster in equivariant architectures while more principal components are required to explain the same variance in non-equivariant ones. Although there is no quantitative difference in the explained variance between the non-equivariant IRC-safe and unsafe architectures, we find that (as one should expect) regression of the leading principal components to a basis of IRC-safe Energy Flow Polynomials (EFPs) performs poorly for the unsafe model. Additionally, while testing on a modified dataset with each jet (i) recoiling against a soft emission outside the jet cone, and (ii) containing an additional soft particle within the jet, we find a considerably large variability in performance as well as per-sample standard deviation across multiple network trainings of the IRC-unsafe model. This variability is smaller for the IRC-safe non-equivariant architecture and decreases further as we impose stronger group equivariance. The E(2) equivariant and IRC-safe architecture remains the most stable.

Our approach offers a concrete method for relating the latent space of deep learning models to the space of analytic, interpretable observables, thereby providing insight into the 'black box' of neural network decision-making. By comparing architectures with varying inductive biases, we elucidate the impact of physics constraints on both performance and interpretability, and highlight the potential of symmetry- and safety-aware networks for robust and insightful applications in collider physics.

The remainder of this paper is organised as follows. In Section~\ref{sec:inductive}, we introduce the notion of strong inductive biases in the context of jet substructure, focusing first on IRC-safe feature extraction and then on equivariance under transformations in the rapidity–azimuth plane, and contrast these with IRC-unsafe constructions. Section~\ref{sec:comparison} provides a comparative analysis of the different bias choices, examining their minimal fibre decompositions, their representability through Energy Flow Polynomials, and their stability in the presence of additional soft emissions. Section~\ref{sec:numeric} details the numerical setup of our study, including the quark–gluon dataset, the graph construction procedure, and the message passing network architectures and training strategies we employ. Section~\ref{sec:results} presents our results in four parts: first, we compare the classification performance of the different networks; second, we investigate the organisation of their latent spaces using principal component analysis; third, we quantify the correspondence between principal components and established jet observables by regressing Energy Flow Polynomials; and fourth, we probe the susceptibility of each model to additional soft radiation. Finally, in Section~\ref{sec:conclusions} we summarise our findings and outline the implications of symmetry- and safety-aware network design for the development of interpretable and reliable machine learning methods in collider physics.

\section{Strong Inductive Biases for Jet Substructure}
\label{sec:inductive}
Deep learning algorithms are highly efficient in extracting relevant features for specific tasks. The physical behaviour of these extracted features can be controlled to some degree via \emph{inductive biases}. Strong inductive biases restrict the function to be equal in a predetermined partition of the domain, thereby restricting the learning process to a representative element of each such partition.  

In this section, we discuss two strong inductive biases relevant to jet substructure: IRC-safe feature extraction, and equivariance in the rapidity-azimuth plane. These biases anchor the learned representation in established jet physics. In the remainder of the paper, we consider all measured particles to be massless, parametrized by $p\equiv (z,y,\phi)$, with $z=p_T/\sum p_T$ being the transverse momentum fraction, $y$ the rapidity, and $\phi$ the azimuthal angle.

\subsection{Infrared and collinear safe feature extraction} 
\label{sec:irc_ext}  
Infrared and collinear safe observables remain unchanged under the addition of arbitrarily soft particles or collinear splitting of a particle. They play a crucial role in the fixed-order calculability of observables in theories containing massless gauge bosons. Such observables form the foundation of rigorous experimental verification of first-principles predictions of perturbative QCD in the high-energy regime. Further nuances due to all-order effects can result in classes of observables requiring stricter criteria like recursive IRC safety~\cite{Banfi:2003je,Banfi:2004yd}, or looser ones like Sudakov safety~\cite{Larkoski:2013paa,Larkoski:2015lea}. Nevertheless, a major portion of the phenomenological program at LHC, particularly for jet substructure, involves constructing IRC-safe observables like energy correlators~\cite{Larkoski:2013eya}, N-subjettiness~\cite{Thaler:2010tr}, etc, and their rigorous understanding~\cite{Dasgupta:2013ihk,Dasgupta:2013via,Larkoski:2017jix}. 

In the framework of energy-weighted message passing~\cite{Konar:2021zdg}, a message passing operation conserves collinearity of the input four vectors $p_i$ if the updated node representations $\mathbf{h}_i$ are the same for two collinear daughters ($r$ and $s$) to its mother ($q$), i.e., for $p_q=p_r+p_s$ and $\hat{p}_q=\hat{p}_r=\hat{p}_s$ we have $\mathbf{h}_q=\mathbf{h}_r=\mathbf{h}_s$. Note that recursive updates of the node representation mimic the collinearity condition, and hence one can compose multiple such collinearity-conserving node updates sequentially. 
As one can expect, this is guaranteed not only by the form of the message passing update but is also dependent on the graph construction algorithm, which must give the same graph in the collinear limit. For instance, a $k$-nearest neighbour graph construction, due to the requirement of a fixed number of neighbours, fails to be structurally collinear safe. Moreover, any algorithm that depends on the magnitude of momentum or energy of constituent particles is collinear unsafe since two collinear daughters can have a continuous-valued energy fraction of the mother. Nevertheless, an essential class of collinear safe graph construction algorithms can be constructed as fixed radius graphs in the rapidity-azimuth plane.
 
For a structurally collinear safe graph, an energy-weighted message passing update of the node features via a trainable message function $f^{(l)}_\theta$ at the $l$-th stage takes the form 
\begin{equation}
\label{eq:empn} 
 	\mathbf{h}^{(l)}_i=\sum_{j\in\mathcal{N}[i]}\; \omega^{(\mathcal{N}[i])}_j \;f^{(l)}_\theta(\mathbf{h}^{(l-1)}_i,\mathbf{h}^{(l-1)}_j)\quad,
\end{equation}
where $\mathcal{N}[i]$ is the fixed-radius neighbourhood of constituent $i$ in $(y, \phi)$ plane, and $\omega^{(\mathcal{N}[i])}_j=p_T^j/\sum_{k\in\mathcal{N}[i]} p_T^k$ are the corresponding momentum fraction weights. This update conserves collinearity in $\mathbf{h}^{(l)}_i$ if it is conserved in $\mathbf{h}^{(l-1)}_i$. Additionally, since the neighbourhood weights $\omega^{(\mathcal{N}[i])}_j$ are momentum fractions, adding an arbitrarily soft particle perturbs the update only at $\mathcal{O}\!\left(p_T^{\text{soft}}\right)$ and vanishes in the strict soft limit. Thus, one can construct IRC-safe extracted graph features by an energy-weighted graph readout on the concatenated node features $\mathbf{H}_i=\bigoplus_{l=1}^{L}\,\mathbf{h}^{(l)}_i$ for up to $L$ successive message passing operations as
\begin{equation}
\label{eq:irc_rep} 
 \mathbf{G}=\sum_{i\in\text{Jet}}\; z_i \; \mathbf{H}_i\quad,
\end{equation}
where $\text{Jet} = \{1, \dots, n\}$ denotes the set of all constituents. Consequently, Eq.~\ref{eq:irc_rep} defines the jet representation used for our analysis.
 
\subsection{Equivariance in the rapidity-azimuth plane} 
Another fundamentally important inductive bias for high-energy physics is group symmetries. More generally termed as equivariance in machine learning and mathematical literature, it subsumes invariance and covariance under a single term. See Appendix~\ref{app:equiv} for the relevant terminology related to group equivariance. In the rapidity-azimuth plane, consider input vectors $\mathbf{x}^{(0)}_i=(\Delta y_{i\small{J}},\Delta \phi_{i\small{J}})$, where the difference is taken with respect to the jet axis. The jet axis $J$ is defined via E-scheme recombination, $P^{\mu}_{\text{Jet}} = \sum_{i=1}^{n} p_i^{\mu}$, yielding $(y_J, \phi_J)$. In this $(y, \phi)$ plane, the Euclidean group E(2) acts by applying the same rotation and translation on all particle positions, whereas O(2) restricts this to rotations only. One can impose energy-weighted, collinearity conserving and E(2)-equivariant message passing operation~\cite{Bhardwaj:2024djv} at the $l$-th message passing stage as
\begin{subequations} 
	\label{eq:irc_en_equiv}
	\begin{align}
		\label{eq:irc_en_equiv_mij}
		\mathbf{m}^{(l)}_{ij}&=\Phi^{(l)}_e(\mathbf{h}^{(l-1)}_i,\mathbf{h}^{(l-1)}_j,|\mathbf{x}^{(l-1)}_i-\mathbf{x}^{(l-1)}_j|^2)\quad,\\
		\label{eq:irc_en_equiv_xi}
		\mathbf{x}_i^{(l)}&=\mathbf{x}^{(l-1)}_i+\sum_{j\in\mathcal{N}[i]} \omega^{(\mathcal{N}[i])}_j\; (\mathbf{x}^{(l-1)}_i-\mathbf{x}^{(l-1)}_j) \; \Phi^{(l)}_x(\mathbf{m}^{(l)}_{ij})\quad,\\
		\label{eq:irc_en_equiv_mi}
		\mathbf{m}^{(l)}_i&=\sum_{j\in\mathcal{N}[i]} \omega^{(\mathcal{N}[i])}_j\;\mathbf{m}^{(l)}_{ij}\quad,\\
		\label{eq:irc_en_equiv_hi}
		\mathbf{h}^{(l)}_i&=\Phi^{(l)}_h(\mathbf{h}^{(l-1)}_i,\mathbf{m}^{(l)}_i)  \quad. 
	\end{align}
\end{subequations}  
Here, $\mathbf{h}^{(.)}_i$ and $\mathbf{x}^{(.)}_i$ are E(2)-scalar and E(2)-vector node features respectively. The maps $\Phi^{(.)}_e$, $\Phi^{(.)}_x$, and $\Phi^{(.)}_h$ are multilayer perceptrons with $\Phi^{(.)}_x$ producing a sigmoid-activated one-dimensional output, used for weighted aggregation.

Similarly, considering $\mathbf{h}^{(.)}_i$ and $\mathbf{x}^{(.)}_i$ as O(2)-scalar and O(2)-vector features, and replacing the updates \eqref{eq:irc_en_equiv_mij}-\eqref{eq:irc_en_equiv_xi} with 
\begin{subequations}
	\label{eq:irc_on_equiv}
	\begin{align}
		\label{eq:irc_on_equiv_mij}  
		\mathbf{m}^{(l+1)}_{ij}&=\Phi^{(l+1)}_e(\mathbf{h}^{(l)}_i,\mathbf{h}^{(l)}_j,|\mathbf{x}^{(l)}_i-\mathbf{x}^{(l)}_j|^2,\mathbf{x}^{(l)}_i\cdot\mathbf{x}^{(l)}_j)\quad,\\
		\label{eq:irc_on_equiv_xi}
		\mathbf{x}_i^{(l+1)}&=\mathbf{x}^{(l)}_i+\sum_{j\in\mathcal{N}[i]} \omega^{(\mathcal{N}[i])}_j\;\mathbf{x}^{(l)}_j \; \Phi^{(l+1)}_x(\mathbf{m}^{(l+1)}_{ij})\quad,
	\end{align} 
\end{subequations} 
we get a collinearity-conserving and O(2)-equivariant message passing update. Note that O(2) being a subgroup of E(2), the E(2)-equivariant updates are also equivariant under O(2) rotations, however Eqs.~\ref{eq:irc_on_equiv} are not E(2)-equivariant by construction since (i) the first expression considers the dot product $\mathbf{x}^{(l)}_i\cdot\mathbf{x}^{(l)}_j$ which is not an E(2)-scalar, and (ii) the second expression is not equivariant under translations $\mathbf{x}_i^{(l)}\mapsto \mathbf{x}_i^{(l)}+\mathbf{t}$, via a common translation vector $\mathbf{t}$ for all $i$. 

Although the intermediate updates are equivariant, we will focus on the physically more relevant invariant graph representations, i.e. we will primarily consider only the scalar part of the equivariant updates for the IRC-safe graph representation as defined in Eq.~\ref{eq:irc_rep}.

\subsection{Comparison with IRC-unsafe feature extraction}
For a fixed multiplicity $n$, let  $\mathcal{J} \subset \mathbb{R}^{3n}$ denote the space of jet configurations, with each jet being treated as a constituent four-vector set\footnote{Taking the jet constituents as a set already imposes an inductive bias via the permutation equivalence relation $\sim_{\sigma}$ on ordered tuples $\mathbf{X}=(p_1,p_2,...,p_n)$, identifying $\mathbf{X}'=(p_{\sigma(1)},p_{\sigma(2)},...,p_{\sigma(n)})\sim_{\sigma}\mathbf{X}$, and $ [\mathbf{X}]_{\sim_\sigma}$ as a single configuration. Since all architectures share this inductive bias, we regard elements of $\mathcal{J}$ to be sets. More appropriately, any inductive bias imposes an additional quotient structure on the domain, with permutation invariance shared by all. Under the quotient topology induced by the Euclidean topology on $\mathbb{R}^{3n}$, one can interchangeably talk of continuous functions on $\mathbb{R}^{3n}$. This unique correspondence arises from the quotient topology's definition (regardless of whether it is metrisable or not) as the \emph{finest} one under which the canonical projection is continuous.} 
$\zeta=\{p_1,p_2,...,p_{n-1},p_n\}$. Let $\zeta_0$ be a jet where a particle $p_n$ is exactly collinear to another particle $p_{i_0}$. Assuming IRC safety, this effectively reduces the multiplicity of such jets to $n-1$ without explicitly considering the intricacies involved with variable-sized sets, which suffices for the present discussion. 

Consider a continuous observable $\mathcal{O}:\mathcal{J}\to\mathcal{Y}$ under the usual Euclidean metric topology. Recall that an observable $\mathcal{O}$ is continuous if the preimage of every open set in $\im(\mathcal{O})\subseteq\mathcal{Y}$ is open in $\mathcal{J}$. If the observable is also IRC-safe, for an open neighbourhood around $\mathcal{O}(\zeta_0)$, its preimage should consist of those jets that differ from $\zeta_0$ by (arbitrarily) soft and collinear emissions.   

Define a set $\mathcal{S}_{\varepsilon,\Delta}(\zeta_0)$ consisting of all those jets obtained from $\zeta_0$ by either (i) relaxing the exact collinearity constraint of $p_n$ by allowing a small but finite angular separation bounded by $\Delta$ with its nearest neighbour, or (ii) having an additional soft particle with transverse momentum bounded by $\varepsilon$. Thus, for any jet $\zeta \in \mathcal{S}_{\varepsilon,\Delta}(\zeta_0)$, 
$$\left( \text{argmin}_{i \in \zeta_0}\Delta R_{in}(\zeta_0) = i_0 \; \text{and} \; 0 \leq \Delta R_{i_0 n}(\zeta) < \Delta \right) \quad \text{or} \quad \; \text{with} \; 0 \leq p_{T,i_0}< \varepsilon \quad , $$ 
where $\Delta R_{ij}^2=(y_i-y_j)^2+(\phi_i-\phi_j)^2$ is the distance between constituents $i$ and $j$ in rapidity-azimuth plane. The parameters $\Delta$ and $\varepsilon$ determine the overall range of values taken by a continuous IRC-safe observable for the different jets in $\mathcal{S}_{\varepsilon,\Delta}(\zeta_0)$, and one can find a neighbourhood of $\mathcal{O}(\zeta_0)$ whose preimage contains $\mathcal{S}_{\varepsilon,\Delta}(\zeta_0)$ as a subset.   

A true target observable $\mathcal{\hat{O}}$ assigns a fixed value to every $\zeta \in \mathcal{J}$, independent of the inductive bias. In practice, with finite (and hence discrete) training data, any deep learning algorithm will (at best) converge to the true value of the target function $\hat{\mathcal{O}}$ on a training set consisting of various jets analogous to $\zeta_0$, and then interpolate over the preimage of open neighbourhoods around each $\hat{\mathcal{O}}(\zeta_0)$. For a continuous IRC-safe observable, the preimage of these open neighbourhoods around $\hat{\mathcal{O}}(\zeta_0)$ will contain the set $\mathcal{S}_{\varepsilon,\Delta}(\zeta_0)$ for finite $\varepsilon$ and $\Delta$. In contrast, if one foregoes IRC safety in a continuous observable, the set $\mathcal{S}_{\varepsilon,\Delta}(\zeta_0)$ need not be contained entirely within the preimage of any open neighbourhood of $\mathcal{O}(\zeta_0)$. For example, as is usually done in various state-of-the-art architectures, using $\log p_T$ in the observable leads to a divergence as $\varepsilon\to0$ unless there is a bounded functional composition within the observable definition. In classification tasks, the boundedness of softmax or sigmoid activation prevents such divergences. Detector specifications also fix a finite lower bound on the measurable momentum, which mitigates such divergences. Nevertheless, for an IRC-unsafe construction, the value of an approximated observable will fluctuate over a large range for different jets in $\mathcal{S}_{\varepsilon,\Delta}(\zeta_0)$ and depend sensitively on specific details of the reconstruction and measurement--an undesirable feature and one of the primary motivations for defining IRC-safe observables.

\section{Comparison of Inductive Biases} 
\label{sec:comparison}
For a jet with massless constituents described by per-particle coordinates $p_i=(z_i,\Delta y_{iJ},\Delta \phi_{iJ})$, where $z_i$ is the transverse momentum fraction, and $(\Delta y_{iJ},\Delta\phi_{iJ})$ are measured relative to the jet axis $J$, consider the jet representation obtained from the following inductive biases:
\begin{enumerate}
	\item $I_E$: E(2)-invariant and IRC-safe 
	\item $I_O$: O(2)-invariant and IRC-safe 
	\item $I_S$: IRC-safe 
	\item $I_U$: IRC-unsafe 
\end{enumerate}
In this section, we discuss the qualitative differences among these four cases on the backdrop of quark/gluon classification. The corresponding network architectures will be described and studied in Section~\ref{sec:numeric}, where we present our numerical analysis. Since our main aim is to get a heuristic understanding of the different biases, we will refrain from a rigorous presentation of the underlying mathematics.

\subsection{Minimal fibre decomposition in the domain}
Consider $\zeta=\{p_1,p_2,...,p_n\}$, the set of jet constituents. We want to look at elements of an equivalence class of jets $[\zeta]_I$ that are guaranteed to have the same representation for an assumed inductive bias $I$. This set will be referred to as the minimal fibres~\cite{Ngairangbam:2024cxc, Maitre:2024hzp} assumed by the strong inductive bias, as they can become larger for non-invertible functions. Denoting the per-particle coordinates as $p_i=(z_i,\mathbf{x}_i)$ with $\mathbf{x}_i=(\Delta y_{iJ},\Delta \phi_{iJ})$, the E(2) group with 2D rotation matrix $\mathbf{R}(\theta)$ and translation vector $\mathbf{t}$ transforms each point as
 \begin{equation}
 \label{eq:joint_action} 
 	(z_i,\mathbf{x}_i)\mapsto (z_i, \mathbf{R}(\theta)\;\mathbf{x}_i+\mathbf{t})\quad. 
 \end{equation}
 The O(2) action is obtained by setting $\mathbf{t}=\mathbf{0}$. Clearly, under translations, all inter-particle distances $\Delta R_{ij}=\sqrt{(y_i-y_j)^2+(\phi_i-\phi_j)^2}$ remain constant, while additionally under pure rotations, the distance from the jet axis $\Delta R_{iJ}$ also remains constant. For $n>2$, angles subtended by two particles at a third particle are also invariant under translations and rotations. However, these angles are determined by the inter-particle distances,\footnote{Intuitively, this follows as a consequence of the side-side-side congruence conditions of triangles formed by three points that have the same distance between each other.} and hence are not independent invariants. 

Group actions of the form of eq.~\ref{eq:joint_action} where a common group element (i.e. the same $\theta$ and $\mathbf{t}$) acts on multiple copies of the base space $\mathcal{X}=\mathbb{R}^2$ (considering the non-trivial action on $\mathbf{x}_i$ as $z_i$ are trivially taken to be scalars) are called joint group actions. Since the Euclidean group action on $\mathbb{R}^2$ is transitive, there are no ordinary invariants. The set of all inter-particle distances $\mathcal{R}_n=\{ \Delta R_{ij}: 1 \le i < j \le n \}$ are joint invariants, and form a fundamental set~\cite{Olver2001}; any other joint invariant is a function of this minimal set. Thus, an E(2) invariant representation will be uniquely parametrised by the inter-particle distances $\Delta R_{ij}$ between any two particles of the set.
 
 In contrast, the O(2) action on $\mathbb{R}^2$ is not transitive; there are individual invariants $\Delta R_{iJ}$ that remain constant under the joint action. The fundamental set of joint invariants for the O(2) group is constructed from the inner product $\mathbf{x}_i\cdot \mathbf{x}_j$. Since $\Delta R_{ij}^2=\Delta R_{iJ}^2+\Delta R_{jJ}^2-2\,\mathbf{x}_i\cdot \mathbf{x}_j$, one can use the set $\mathcal{R}_n$ along with the individual invariants to characterise an O(2) joint orbit. Consequently, two jets that have the same E(2) jet representation will have the same O(2) representation if, in addition, all particles have the same distance $\Delta R_{iJ}$ from the jet centre.

\subsection{Representability via Energy Flow Polynomials}
 Energy Flow Polynomials (EFPs)~\cite{Komiske:2017aww} are a class of observables that form an over-complete basis in the space of all IRC-safe observables. An EFP is defined through an unlabelled, undirected multigraph $G$ having $N$ nodes as   
 \begin{equation}
 	\text{EFP}_{G}=\sum_{i_1}...\sum_{i_N}\;z_{i_1}\;z_{i_2}...\;z_{i_N} \prod_{(i,j)\in G} \Delta R_{ij} \quad,
 \end{equation}
 where the sum runs over all particles. Because EFPs depend only on the relative position $\Delta R_{ij}$ by construction, they naturally exclude the class of IRC-safe observables that depend on the jet's overall direction, like generalised angularities. Such observables, which explicitly depend on the jet axis, are generally susceptible to recoil effects. Moreover, when they depend on the jet rapidity, such observables break longitudinal boost invariance. 

Consider the approximation of IRC-safe observables in terms of C-correlators~\cite{Tkachov:1995kk} 
 \begin{equation}
 	\mathcal{O}(J)\approx\sum_{N=0}^{N_{\text{max}}}\; \mathcal{C}_{N}^{f_N} \quad,\quad \mathcal{C}_N^{f_N}=\sum_{i_1,i_2,...i_N} z_{i_1}\,z_{i_2},...,z_{i_N}\;f_N(\mathbf{x}_1,\mathbf{x}_2,...,\mathbf{x}_N)\quad,     
 \end{equation} 
 where $f_N$ are permutation-invariant angular weighting functions reflecting the indistinguishability of jet constituents. The imposition of group invariance imposes restrictions on the explicit dependence of angular weighting functions $f_N$ by making them depend on group invariants alone. As a consequence of transitivity of the E(2) action, E(2) invariant functions are independent of individual positions $\mathbf{x}_i$, so $f_1$ can at most be constant $$ \text{E(2)-invariance}\implies f_1(\mathbf{x}_i)=k\implies\mathcal{C}_{1}^{f_1}=k\sum_{i}\;z_i \quad,$$where $k$ is some real constant. On the other hand, for O(2) invariant functions $f_1$ may depend on $\Delta R_{iJ}$, i.e. $$\text{O(2)-invariance}\implies f_1(\mathbf{x}_i)=f(\Delta R_{iJ})\implies\mathcal{C}_1^{f_1}=\sum_i z_i\;f_1(\Delta R_{iJ})\quad.$$

We can now write the general form of the angular weighting functions for any $N\geq 2$ as 
\begin{equation} 
    \begin{split} 
    &\text{E(2)-invariance}\implies f_N(\mathbf{x}_1,\mathbf{x}_2,...,\mathbf{x}_N)=f_N(\mathcal{R}_N)\quad,\\ 
    &\text{O(2)-invariance}\implies f_N(\mathbf{x}_1,\mathbf{x}_2,...,\mathbf{x}_N)=f_N(\Delta R_{i_1J},\Delta R_{i_2J},...,\Delta R_{i_NJ},\mathcal{R}_N)\quad .
    \end{split} 
\end{equation}
 
Clearly, the class of observables represented by EFPs is E(2)-invariant. Since O(2) invariant features may depend on $\Delta R_{iJ}$, it is not guaranteed that they will be well-represented by EFPs. Moreover, for non-equivariant IRC-safe feature extraction, one expects EFPs to perform even worse than O(2) invariant features since no group structure constrains them. Even worse would be IRC-unsafe quantities, as these features lie beyond the representational capacity of EFPs. Thus, while the O(2)-invariant and non-invariant IRC-safe features can lie in the space of functions represented by EFPs, IRC-unsafe ones will always lie outside their span.

\subsection{Impact of soft emissions} 
Before discussing the effects of emissions for arbitrary multiplicities, we discuss a simple 2-particle case to build up the intuition. Consider a jet $\zeta=\{p,q\}$ with two resolved particles with $p=(z, y,\phi)$ and $q=(s,y_s,\phi_s)$ with $z \gg s$, and another jet $\zeta'=\{p,q'\}$ with $q'=(s,y'_s,\phi'_s)$. For any continuous IRC-safe observable $\mathcal{O}$ defined on $\zeta$ and $\zeta'$, we expect the observable values to be close due to the softness of $q$ or $q'$. However, additional assumptions on the angular coordinates will determine the degree of stability. Here, the degree of stability means the robustness of the observable when the coordinates $(y_s,\phi_s)$ and $(y'_s,\phi'_s)$ are varied. Already, one can expect that the assumption of symmetries in the rapidity-azimuth plane will stabilise the observable $\mathcal{O}$ over a larger range as long as both $\zeta$ and $\zeta'$ lie close in the orbit space. 

For $I_S$ and $I_U$, the observable is guaranteed to remain stable only when $(y'_s,\phi'_s)$ is restricted to a small neighbourhood around $(y_s,\phi_s)$, since there are no additional symmetries. Moreover, IRC-unsafe feature extraction does not guarantee $\mathcal{O}(\zeta)$ or $\mathcal{O}(\zeta')$'s closeness to $\mathcal{O}(\zeta_0)$, where $\zeta_0$ is a jet without the additional resolved emission, (i.e $\zeta_0=\{p,k\}$ with $k$ collinear to $p$), and is already pathological from the outset. For O(2) invariance, the observable remains stable for those jets $\zeta'$ that can be rotated, so that $(y'_s,\phi'_s)$ lies in the neighbourhood of $(y_s,\phi_s)$. Geometrically, this corresponds to a ring of some finite but small width around $(y,\phi)$. An E(2) invariant observable will be stable for those jets $\zeta'$ that have the angular separation $\Delta R_{pq'}$ in the neighbourhood of $\Delta R_{pq}$. This consists of all possible rings constructed out of two circles having a slight difference in radius around $\Delta R_{pq}/2$ sharing the same origin anywhere in the plane. Even though the two particles can be translated anywhere or rotated around the translated jet axis from the perspective of E(2) invariance, the kinematics between the two particles determine the allowed range. 

The situation remains similar if one increases the number of constituents. Any jet $\zeta'$ will be stable if its particles have similarly valued $z_i$, and $(y'_i,\phi'_i)$ lies around a small neighbourhood of $(y_i,\phi_i)$ of the particle in the jet $\zeta$ for a learned $I_S$-observable.  Such jets $\zeta'$ will be \emph{almost superimposed} with $\zeta$ in the rapidity-azimuth plane. For an $I_O$-observables, it will be stable for all those jets $\zeta'$ that can be practically superimposed with $\zeta$ via rotations around the jet centre, while for $I_E$-observables, the range of jets will be those that can be almost superimposed by a combination of translations and rotations (around the translated jet axis), the allowed operations determined by kinematic constraints. For an $I_U$-observable, the situation will be similar to that of $I_S$  for the angular coordinates, but will be highly dependent on the soft particles. Note that one can have a similar stability analysis (albeit with larger variation) if one considers the collimated hard subjets as individual particles.

\section{Numerical Analysis}
\label{sec:numeric}
This section presents the numerical framework of our analysis for quark–gluon discrimination using graph neural networks. We describe the dataset, the graph-based jet representation, the network architectures, and the training procedure that form the basis of our study.

 \subsection{Dataset}
The dataset used in this work for quark-gluon tagging is the publicly available dataset~\cite{komiske_2019_3164691}. It is based on proton-proton collisions generated with \texttt{Pythia 8.235}~\cite{Sjostrand:2014zea} at centre-of-mass energy $\sqrt{s} = 14 ~ \text{TeV}$. Quark jets are produced from \texttt{WeakBosonAndParton:qg2gmZq}, and gluon jets from \texttt{WeakBosonAndParton:qqbar2gmZg} with the $Z$ boson decaying to neutrinos. Jets are required to have transverse momentum $p_{T} \in [500, 550]\,\text{GeV}$ and rapidity $|y| < 1.7$. They are clustered with \texttt{FastJet} 3.3.0~\cite{Cacciari:2011ma} using the anti-$k_{T}$ algorithm~\cite{Cacciari:2008gp} with radius parameter $R = 0.4$. The dataset contains 2 million jets, equally split between quarks and gluons, restricted to only light quark flavours. Each jet is represented as an array of its particle constituents with features $(p_T, y, \phi, \mathrm{PID})$. All jets have an accompanying binary label, with gluons assigned 0 and quarks assigned 1. We divide the dataset into training, validation, and testing sets in the ratio $(0.6, 0.2, 0.2)$. No detector effects or pileup are applied. 

\paragraph{Input representation}
Each jet is represented as a graph whose nodes correspond to jet constituents. For a given jet, all constituents are treated as massless and described by their transverse momentum fraction $z_i = p_{T,i}/\sum_{j \in \text{Jet}} p_{T,j}$ and angular coordinates $(y_i, \phi_i)$. The jet axis $J$ is taken to be the direction of the jet four-momentum, defined as the sum of the constituent four-momenta. To construct the input graphs, we use $\Delta R_{ij}$ to construct a radius graph connecting all pairs of particles with $\Delta R_{ij} < r_{0} = 0.5$. Using a radius graph ensures that the graph topology remains well-behaved in the soft and collinear limits, which is essential for maintaining IRC safety of the representation.

The node features consist of both vector and scalar components, while $z_i$ serve as node weights in the IRC-safe aggregation operation. The vector features are the coordinates of each particle with respect to the jet axis,
\begin{equation}
\mathbf{x}^{(0)}_{i} = (\Delta y_{iJ}, \, \Delta \phi_{iJ}).
\end{equation}
The radial distance of the particle from the jet axis is used as the scalar node representation 
\begin{equation}
\mathbf{h}^{(0)}_{i} = \Delta R_{iJ},
\end{equation}
for the O(2) equivariant model.
For each edge $j\mapsto i$, we evaluate the edge weight  $\omega^{(\mathcal{N}[i])}_j$ constructed out of the transverse momentum fraction as $\omega^{(\mathcal{N}[i])}_j=p_T^j/\sum_{k\in\mathcal{N}[i]} p_T^k$. This weighting scheme reflects the physical intuition that harder neighbours should contribute more strongly in the message passing operation.

The full jet graph is stored in a \texttt{torch\_geometric.data.Data} object containing the edge indices, scalar and vector node features, node weights, and edge weights, along with the jet label. This provides the input representation for the GNNs.

\subsection{Network architectures and training}
\begin{figure}[t]
\centering
\includegraphics[scale=0.51]{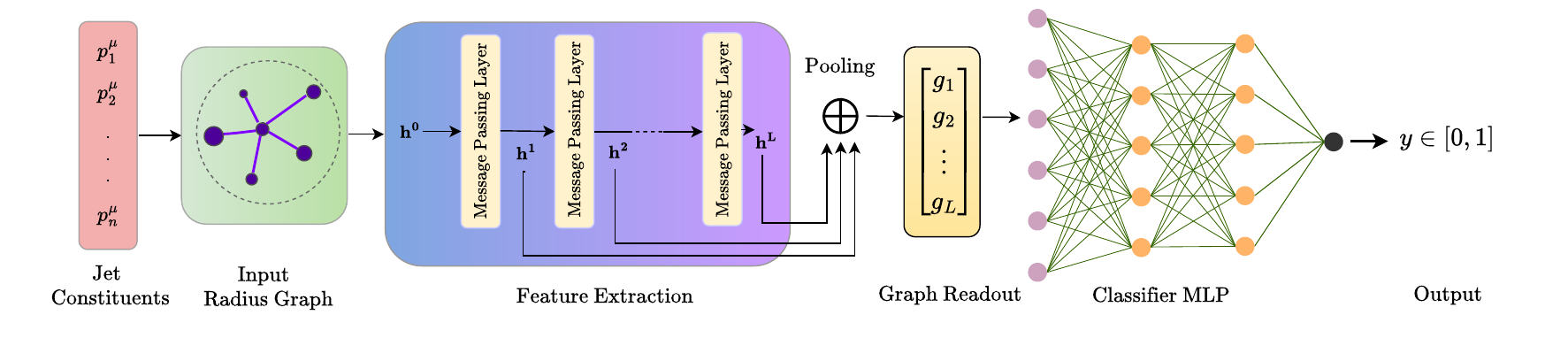}
\caption{Schematic diagram showing the common network architecture used for all models.}
\label{fig:network}
\end{figure}

\paragraph{Network architectures}
All networks share a standard structure consisting of multiple message passing layers, followed by a graph readout and a fully connected multi-layer perceptron (MLP) for classification (see Figure~\ref{fig:network}). The message passing layers update the node features by exchanging information between connected particles, while the readout operation aggregates these into a single jet-level latent vector. The final MLP maps this latent representation to a one-dimensional logit score for quark-gluon discrimination. 

The inductive biases introduced in Section~\ref{sec:comparison} are realised in practice through four graph neural network models. The models corresponding to inductive bias categories $I_E$ and $I_O$ are implemented as E(2)-EMPN and O(2)-EMPN respectively. These are E(2) and O(2) equivariant energy weighted message passing networks that are IRC-safe by construction. The $I_S$ and $I_U$ cases are instantiated as EMPN (energy-weighted message passing, IRC-safe) and MPNN (message passing neural network, IRC-unsafe), both employing EdgeConv-style message passing.

The structure of equivariant networks is defined in Eqs. \ref{eq:irc_en_equiv}, where $\mathbf{x}^{(.)}_i$ denote the vector features, $\mathbf{h}^{(.)}_i$ the scalar node features, and $\mathbf{m}^{(.)}_{ij}$ the intermediate message. The learnable functions $\Phi^{(.)}_e$, $\Phi^{(.)}_x$, and $\Phi^{(.)}_h$ are MLPs with two hidden layers of 128 units each and ReLU activation. $\Phi^{(.)}_x$ outputs a one-dimensional coefficient with Sigmoid activation, while $\Phi^{(.)}_e$ and $\Phi^{(.)}_h$ output 128-dimensional embeddings with Linear activation. 

The EdgeConv models use a simpler message passing scheme with a single MLP per layer. This MLP $\Phi$ also contains two hidden layers with 128 units with ReLU activation, and outputs a 128-dimensional embedding. In the IRC-safe variant (EMPN), the inputs are node coordinates, and the message passing is again energy-weighted to preserve infrared and collinear safety. In the IRC-unsafe MPNN, no such modifications are applied. The inputs are the raw particle features\footnote{To keep the comparison fair, the MPNN is trained using only the same four-momentum information as the other networks, without additional IRC-unsafe inputs.} $(z, y, \phi)$  of each connected pair, which can introduce sensitivity to IRC-unsafe effects.

All networks have three message passing layers. Each message passing step produces 128-dimensional scalar features per node. Concatenating the output from all three layers yields a representation of shape $(n, 384)$, where $n$ is the number of jet constituents. Then, for IRC-safe networks, we take an energy-weighted graph readout,
\begin{equation}
\label{eq:graph_readout}
\mathbf{G} = \sum_{i \in \text{Jet}} z_i\; (\mathbf{h}^{(1)}_i\oplus\mathbf{h}^{(2)}_i\oplus\mathbf{h}^{(3)}_i)\quad,
\end{equation}
where $z_i$ are the transverse momentum fractions of each node. For the IRC-unsafe MPNN, the readout is a simple unweighted mean over $\mathbf{h}_i$. In each case, after pooling, we have a 384-dimensional latent jet-embedding vector. This graph readout feature $\mathbf{G}$ is saved for further analysis. It is then passed to the classifier head, an MLP with input dimension 384, two hidden layers of 128 units with ReLU activations, and a final output node producing a logit score.

The equivariant networks E(2)-EMPN and O(2)-EMPN contain approximately 430k trainable parameters, while the EdgeConv-based models, EMPN and MPNN, contain 230k. The larger size of the equivariant models comes from each message passing step using three $\Phi$ networks, in contrast to a single one in the EdgeConv models.

\paragraph{Training}
The dataset comprises 1.2 million training samples and 0.4 million samples each for validation and testing. All networks produce a one-dimensional logit score, and training is performed for quark-gluon discrimination using the binary cross-entropy loss with logits. Models are trained for 75 epochs with a batch size of 128. The optimiser used is \textsc{Adam} \cite{adamopt} with an initial learning rate of $10^{-3}$, which is reduced by a factor of $0.5$ if the validation loss does not improve for three consecutive epochs. The model with the lowest validation loss is retained for inference. 
All experiments are repeated 10 times with random initialisation. 
The networks are implemented in \texttt{PyTorch} 2.2.2~\cite{pytorch} with \texttt{PyTorch Geometric} 2.5.2,~\cite{fey2025pyg}, and training was performed on NVIDIA A100 GPUs.

\section{Results} 
\label{sec:results}
This section reports our main findings. We begin by comparing quark/gluon classification across the four architectures. We then examine how each model organises its latent space via principal component analysis and relate the leading components to Energy Flow Polynomials through linear regression. Finally, we test the robustness of these models under controlled soft-emission perturbations.

\subsection{Classification performance}
\label{subsec:classification_performance}

The numerical results for classification are summarised in Table \ref{tab:network_results}, which reports the area under the ROC curve (AUC) and the accuracy for each network. As we can see, all networks achieve competitive performance with mean AUC values $\sim 0.89 - 0.90$ and accuracies $\sim 0.825$, with standard deviations at the level of order $10^{-4} - 10^{-3}$. Importantly, the IRC-safe models (E(2)-EMPN, O(2)-EMPN, EMPN) perform on par with the unconstrained and IRC-unsafe MPNN. This establishes that incorporating physical constraints such as equivariance and IRC safety does not degrade classification ability.\footnote{We expect the performance of MPNN can be further improved with the incorporating other IRC-unsafe input features, such as $\mathrm{PID}$, etc.} Instead, the parity in performance indicates that an IRC-safe combination of features fundamentally captures the discriminating information between quark and gluon jets. Indeed, the optimal quark-gluon likelihood ratio is an IRC-safe observable~\cite{Larkoski:2019nwj}, which explains why constraining networks to this regime still allows them to reach maximal discrimination.

\begin{table}[t]
	\centering
	\begin{tabular}{lccc}
		\hline
		\textbf{Network} & \textbf{IRC-Safe} & \textbf{AUC} & \textbf{Accuracy} \\
		\hline
		E(2)-EMPN    & Yes & 0.8992 $\pm$ 0.0004 & 0.8254 $\pm$ 0.0005 \\
		O(2)-EMPN    & Yes & 0.8988 $\pm$ 0.0005 & 0.8249 $\pm$ 0.0007 \\
		EMPN         & Yes & 0.9000 $\pm$ 0.0005 & 0.8262 $\pm$ 0.0007 \\
		MPNN         & No  & 0.8985 $\pm$ 0.0003 & 0.8247 $\pm$ 0.0003 \\
		\hline
	\end{tabular}
	\caption{AUC and Accuracy for different networks on the quark vs gluon classification task. Values are quoted as mean $\pm$ standard deviation over 10 training instances.}
	\label{tab:network_results}
\end{table} 

From a physics standpoint, this outcome is promising. The fact that networks constrained to respect physically meaningful symmetries converge to the same level of performance as the MPNN demonstrates that the essential quark-gluon differences reside in robust, symmetry-respecting features. Thus, we can impose physical structure on the networks without paying any penalty, and in doing so, we gain interpretability and confidence that the learned decision boundaries are grounded in meaningful physics. Since all networks reach the same level of separation, the next step is to examine what information they rely on to attain this performance. As we will see in Section \ref{subsec:soft_emission_sensitivity}, different models may arrive at different decision boundaries despite matching the global performance, and interesting differences emerge when we test these models under controlled modifications of the input jets.

\subsection{Principal components of the latent space} 

To better understand how each model internally represents data, we analyse the structure of their latent space. A standard method for analysing high-dimensional representations is principal component analysis (PCA), a linear dimensionality method that identifies orthogonal directions of maximum variance. PCA provides a ranked basis of axes onto which data can be projected, revealing dominant modes of variation. Further mathematical details of PCA are provided in Appendix~\ref{app:pca}. 

For this study, we apply PCA to the graph-level embedding $\mathbf{G}$, as mentioned in Eq.~\ref{eq:graph_readout}. This graph-readout corresponds to the concatenated outputs of all message passing layers, which results in a 384-dimensional\footnote{Three message passing layers with 128-dimensional outputs each.} feature vector per jet. These graph-readout features are further passed to a downstream classifier for the discrimination task. We store these embeddings for all 10 independent runs of each network. PCA was applied to these graph-readout latent features to study their structure. 

\begin{figure}[t]
    \centering
    \includegraphics[scale=0.4]{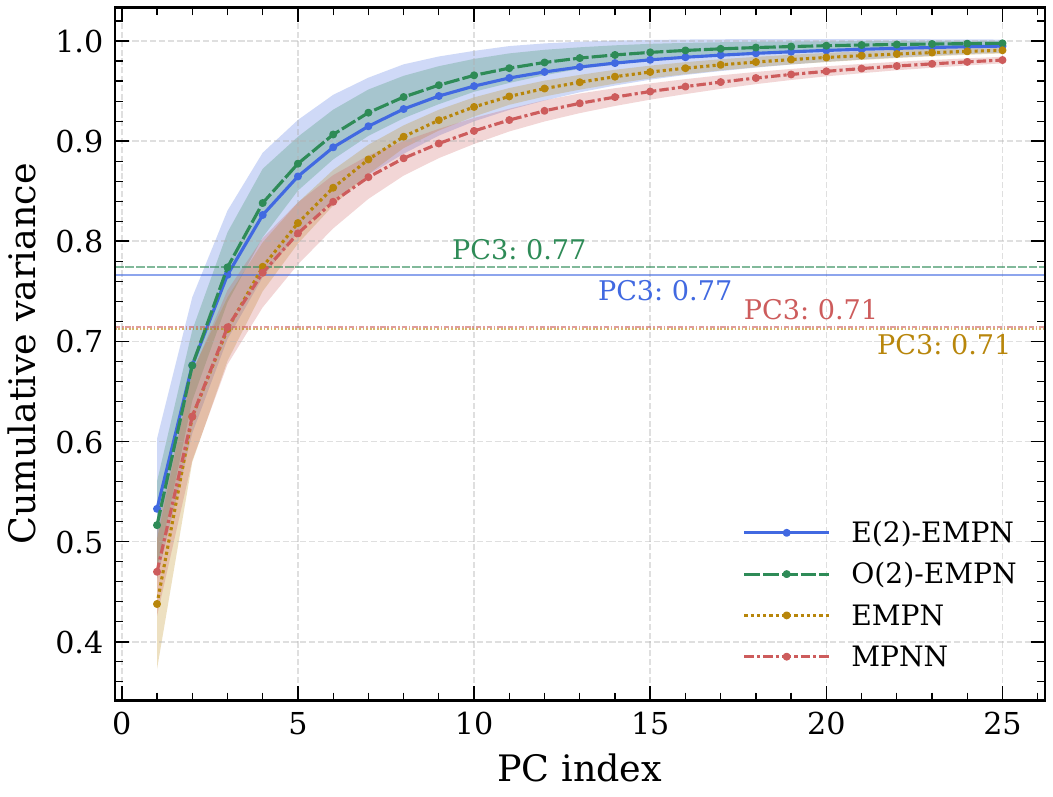}
    \includegraphics[scale=0.4]{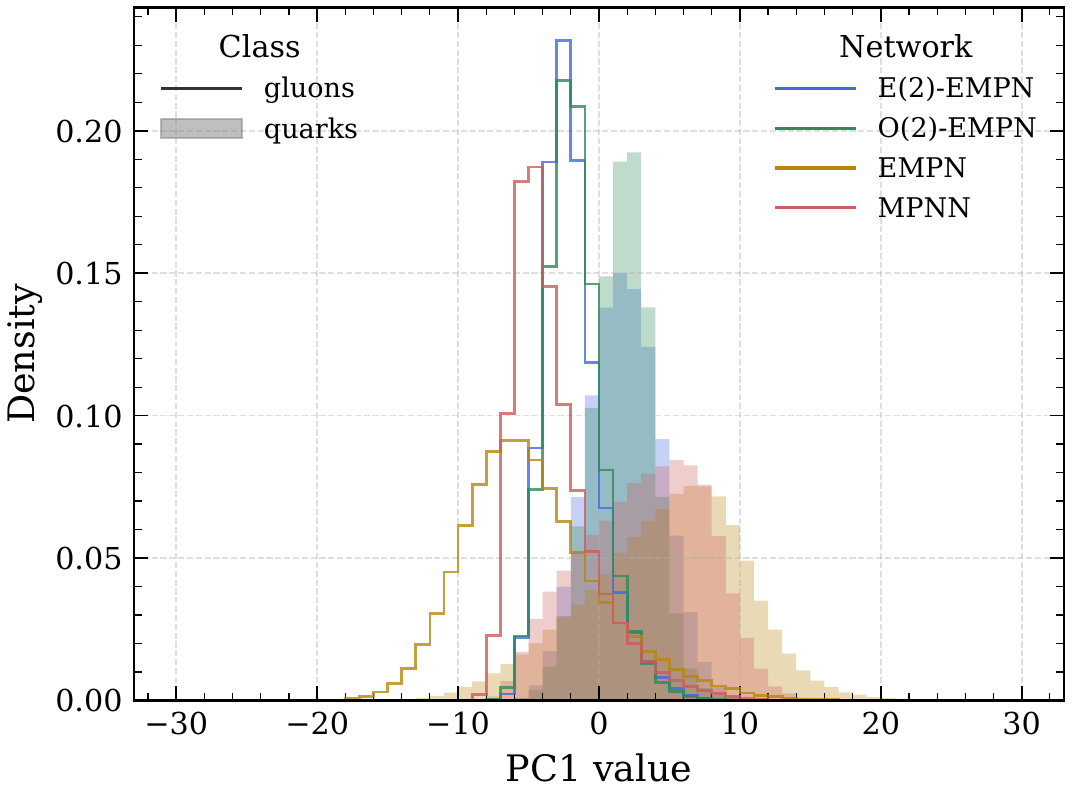}
    \caption{Left: Cumulative explained variance ratio per principal component (PC) of the graph readout features across networks. The curves show the mean across 10 runs, and the shaded bands indicate one standard deviation. Right: Distribution of the first principal component (PC1), averaged over 10 runs, for quark and gluon jets shown separately across networks.}
    
    \label{fig:pca_latent}
\end{figure}

The left plot in Figure \ref{fig:pca_latent} shows the cumulative variance explained by the leading principal components, where the curves show the mean across runs, and the shaded band indicates one standard deviation. As we see, the variance saturates near unity after only a few components, confirming that the effective dimensionality of the latent jet representation is far smaller than the 384-dimensional raw feature space. This rapid saturation is consistent with the manifold hypothesis, which states that high-dimensional data often lie on a much lower-dimensional submanifold.

The cumulative variance curves highlight a clear distinction in the latent space structure between equivariant and non-equivariant models. For the E(2)-EMPN and O(2)-EMPN architecture, the first three PCs capture about $77\%$ of the variance, while for EMPN and MPNN, they capture only about $71\%$. This indicates that the equivariant models concentrate their jet representation more strongly into the leading directions, whereas the non-equivariant models distribute information slightly more broadly across subleading axes. Physically, this reflects the more substantial inductive bias of the equivariant networks. By enforcing symmetry and IRC safety, they compress the representation into fewer dominant axes that align with symmetry-respecting features.
In contrast, the non-equivariant models leave more variance in the subleading PCs, possibly reflecting additional flexibility to capture residual, less-symmetric aspects of the data. Significantly, none of the models have collapsed the variance into a single latent direction, but they retain multiple independent features relevant for the complex quark/gluon discrimination task. This multi-faceted internal representation is encouraging for interpretability; it suggests that these models might be separately encoding different known physical discriminants rather than entangling everything into an inscrutable combination.

To further characterise the latent space, the class-wise distribution of each network's leading principal component (PC1) is shown on the right in Figure \ref{fig:pca_latent}. Across all models, PC1 already exhibits visible class separation. However, the distributions differ in their overall spread, with the equivariant networks showing relatively narrow ranges, the MPNN extending somewhat broader, and the EMPN displaying the widest spread. This indicates that while PC1 encodes the dominant direction across all architectures, each model emphasises slightly different aspects of jet structure along this axis. 

\begin{table}[t]
	\centering
	\begin{tabular}{l|ccc}
		\toprule
		\textbf{Network} & \multicolumn{3}{c}{\textbf{Graph Readout}} \\
		& PC1   & PC2   & PC3    \\
		\midrule
		E(2)-EMPN       & 0.8731 & 0.5312 & 0.5321  \\
		O(2)-EMPN       & 0.8763 & 0.5102 & 0.5025  \\
		EMPN            & 0.8883 & 0.5487 & 0.5013  \\
		MPNN            & 0.8874 & 0.6051 & 0.5051  \\
		\bottomrule
	\end{tabular}
	\caption{AUC scores for the first three averaged principal components (PC1-PC3) of the graph readout feature across all networks. Each value corresponds to the classification performance obtained when using a single PC as a one-dimensional score to separate quark and gluon jets.}
	\label{tab:auc_pcs}
\end{table}

To probe the discriminative structure of the latent space, we examine how well the PCs of graph-readout features, averaged over 10 runs, separate quark from gluon jets. We take the top three PCs and calculate the AUC using each PC as a one-dimensional classification score. The results are shown in Table \ref{tab:auc_pcs}. 

We find that the dominant direction (PC1) is strongly aligned with the classification task for all networks. PC1 achieves AUCs in the range $0.87-0.89$, nearly saturating the full network performance. The fact that averaged PCs retain this level of discriminative power indicates that the networks consistently orient their dominant direction in a similar way across runs. In other words, the most significant axis of variation in the latent space correlates strongly with IRC-safe physics, corroborating the likelihood ratio's IRC safety.

The picture is different for subleading components. For the equivariant and IRC-safe models, PC2 and PC3 have AUCs close to $0.5$, indicating that they carry little classification power on their own. The MPNN, however, shows a relatively better performance with PC2 (AUC $\sim 0.62$), suggesting that the secondary directions also carry useful class information. This points to a subtle difference in how the networks organise their latent space. While all models perform similarly and align their dominant axis with quark-gluon separability, the unconstrained MPNN appears free to distribute part of that information into additional directions. In contrast, constrained architectures rely more on a combination of weaker but symmetry-respecting components, which are tied to physically motivated features. 

To assess this more directly, we next regress the PCs against known IRC-safe observables. From an interpretability perspective, this allows us to test whether the information captured by the PCs corresponds to known, physically meaningful observables.  

\subsection{EFP representation of the latent space}
\label{subsec:efp_regression}

To establish a concrete connection between the learned latent features of the networks and known analytic observables, we fit a basis of Energy Flow Polynomials (EFPs) onto the leading principal components of the graph readout representations. We perform linear regression\footnote{Ordinary least squares linear regression was implemented in \texttt{scikit-learn}'s~\cite{scikit-learn} \texttt{LinearRegression} class.} on the top three PCs of the test dataset, averaged over 10 runs, against all EFPs with degree (number of edges) $d \leq 7$, which consists of $\sim 1000$ composite multigraphs. These are $n$-point correlators, which linearly span the space of IRC-safe observables. This provides a quantitative test of whether the most prominent PCs can be reconstructed from known QCD-inspired quantities.

\begin{figure}[t]
	\centering
	\includegraphics[scale=0.4]{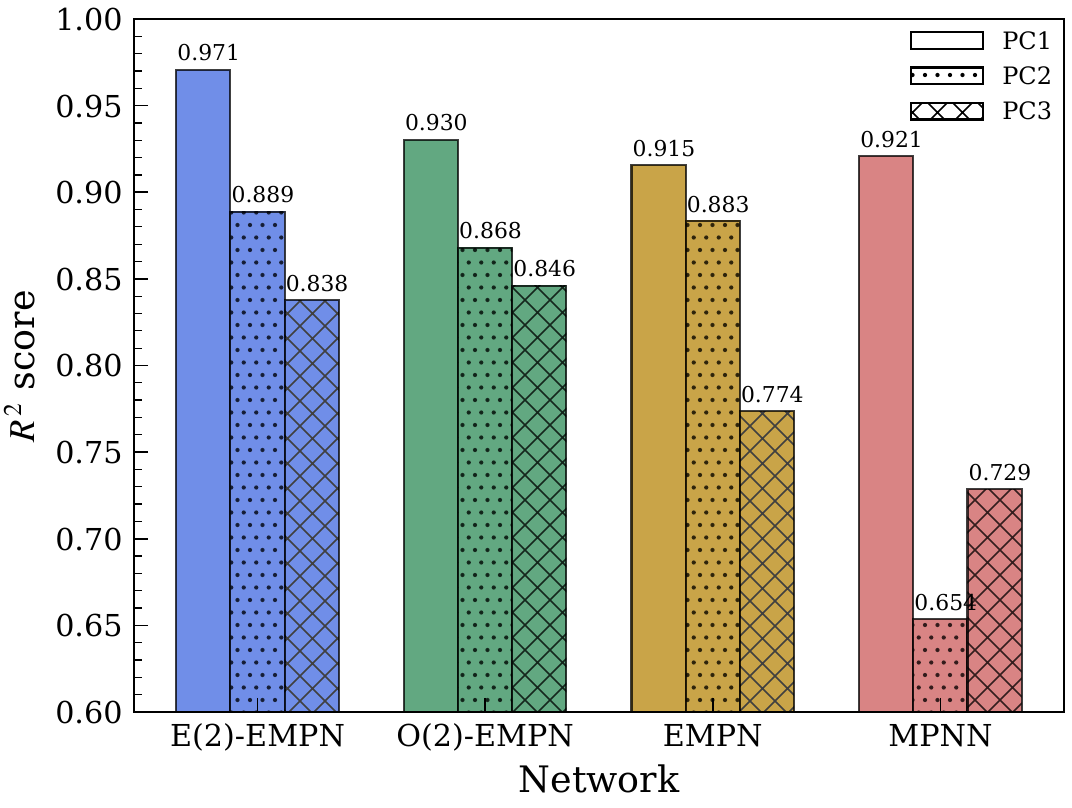}
	\caption{$R^2$ score for regression of EFPs onto the top three averaged principal components of the graph readout feature across networks. The score quantifies the amount of variance in the PCA-reduced features captured by EFPs.}
	\label{fig:r2_bar}
\end{figure}

The regression results are summarised in Table \ref{tab:regression_result}. We report three complementary metrics. The mean squared error (MSE) is a scale-dependent quantity and reflects the absolute reconstruction error. The coefficient of determination ($R^2$) measures the fraction of variance in the PC values explained by the EFP fit, defined as
\begin{equation} 
R^2 = 1 - \frac{\sum_i (y_i - \hat{y}_i)^2}{\sum_i (y_i - \bar{y})^2},
\end{equation}
where $y_i$ are the PC values and $\hat{y}_i$ are the regression predictions. A perfect linear fit gives $R^2=1$. Finally, the mean absolute error normalised by the PC's standard deviation (MAE$/\sigma$) removes sensitivity to overall scale and allows direct comparison across PCs and networks. Among these, $R^2$ provides the most direct measure of how faithfully a latent direction can be expressed in terms of the EFP basis. Therefore, we also display the $R^2$ values obtained for the top three PCs across all networks in Figure \ref{fig:r2_bar}.

\begin{table}[t]
	\centering
	\renewcommand{\arraystretch}{1.3}
	\begin{tabular}{l l | ccc }
		\toprule
		\multirow{2}{*}{\textbf{Network}} & \multirow{2}{*}{\textbf{PC}} 
		& \multicolumn{3}{c}{\textbf{Graph Readout}} \\
		\cmidrule(lr){3-5}
		& & \textbf{MSE} & \textbf{R$^2$} & \textbf{MAE/$\sigma$} \\
		\midrule
		\multirow{3}{*}{E(2)-EMPN}& PC1 & 0.2533 & 0.9708 & 0.1351  \\
		& PC2 & 0.2630 & 0.8888 & 0.2565 \\
		& PC3 & 0.0508 & 0.8376 & 0.2991\\
		\midrule
		\multirow{3}{*}{O(2)-EMPN} 
		& PC1 & 0.4958 & 0.9303 & 0.2034 \\
		& PC2 & 0.5867 & 0.8679 & 0.2740  \\
		& PC3 & 0.1702 & 0.8460 & 0.2944 \\
		\midrule
		\multirow{3}{*}{EMPN} 
		& PC1 & 4.2226 & 0.9155 & 0.2231 \\
		& PC2 & 2.5623 & 0.8834 & 0.2575 \\
		& PC3 & 1.0994 & 0.7737 & 0.3420 \\
		\midrule
		\multirow{3}{*}{MPNN} 
		& PC1 & 1.4403 & 0.9207 & 0.1873  \\
		& PC2 & 0.3745 & 0.6536 & 0.4060  \\
		& PC3 & 0.2107 & 0.7287 & 0.3631  \\
		\bottomrule
	\end{tabular}
	\caption{Results for linear regression of the top three averaged principal components (PC1-PC3) of the graph readout feature onto the EFP basis ($d \leq 7$). Metrics reported are mean squared error (MSE), coefficient of determination ($R^2$), and mean absolute error normalised by the PC standard deviation (MAE$/\sigma$).}
	\label{tab:regression_result}
\end{table}

For the equivariant and IRC-safe models, the fits are consistently strong. PC1 of the E(2)-EMPN network is almost perfectly reconstructed with $R^2 = 0.97$. The O(2)-EMPN and EMPN are not far behind with $R^2 = 0.93$ and $0.92$, respectively. The subleading components, PC2 and PC3, also remain well aligned with high $R^2$ values, in the $0.77$ - $0.89$ range. This demonstrates that multiple latent directions, not just the leading one, can be expressed as a linear combination of EFPs; i.e. the networks constrained with physical inductive bias structure their internal representation around known QCD observables. In contrast, the unconstrained MPNN yields a comparable PC1 fit with $R^2 = 0.92$, but for PC2 and PC3 it degrades sharply ($0.65$ and $0.73$). This suggests that while the leading direction overlaps substantially with IRC-safe observables, a significant fraction of the subleading structure lies outside this EFP basis. 

The error metrics reinforce this picture. For IRC-safe architectures, MSE and MAE/$\sigma$ remain low across PCs, showing stable reconstruction. Meanwhile, for MPNN, PC1 errors are small, but PC2 and PC3 show larger normalised errors. This distinction is meaningful, even though the overall PCA variance structure and the classification performance are similar across networks, the way information is organised internally is different. The constrained models distribute their features into several axes that can be traced back to IRC-safe observables. At the same time, the unconstrained MPNN concentrates useful information in PC1 and leaves the rest in directions that are harder to interpret. The good alignment of PC1 across all networks is expected, since the quark-gluon likelihood ratio itself is IRC-safe; however, the differences in subleading components show how architectural choices affect what additional features are learned. 

\begin{figure}[t]
	\centering
        \includegraphics[scale=0.75]{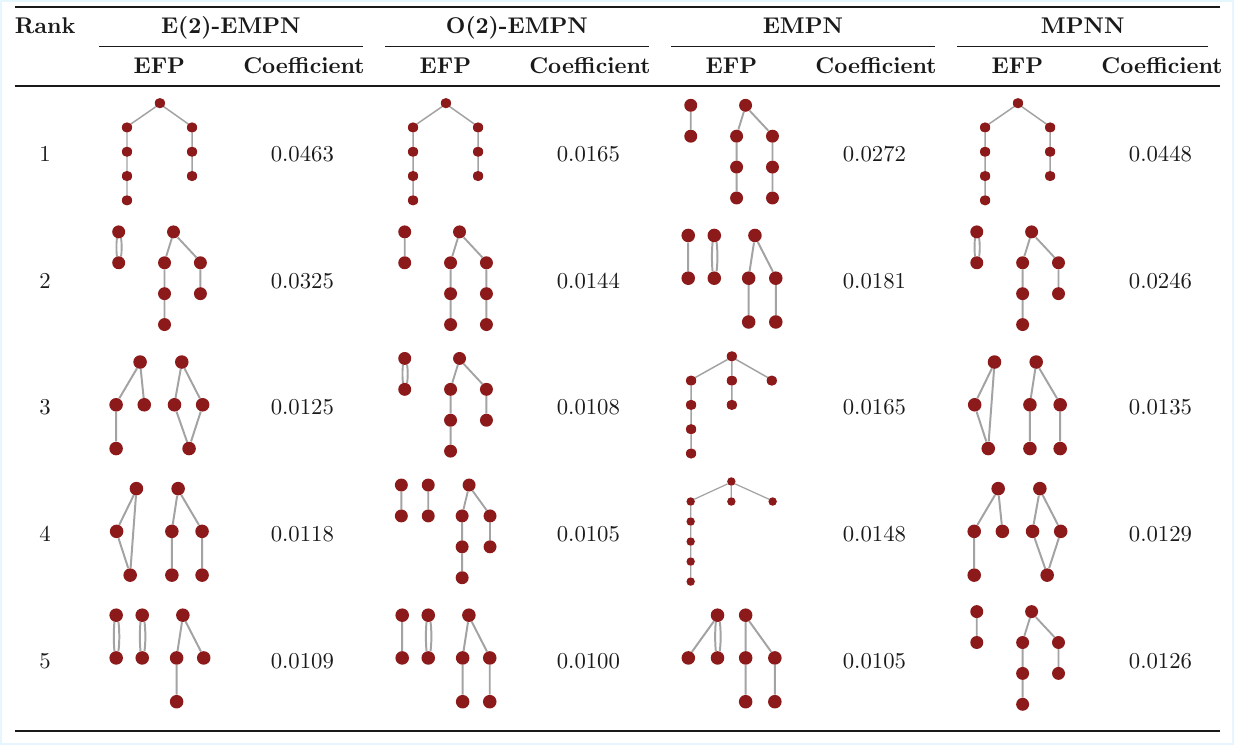}
	\caption{Top-ranked EFPs for PC1 regression across networks, ranked by the magnitude of the standardised coefficient.}
	\label{fig:efp_graphs}
\end{figure}

Beyond overall fit quality, this motivates a closer examination of which specific EFPs dominate the fits and what aspects of jet structure they probe. Figure \ref{fig:efp_graphs} shows the top five EFPs contributing to PC1 for each network, ranked by the magnitude of their standardised regression coefficient. Standardisation rescales the EFPs to unit variance, so a coefficient measures the effect of one $\sigma$ change in that observable. This removes the biases resulting from the differing numerical ranges of the EFPs and allows for a meaningful comparison of relative importance. For comparison, we additionally normalise the magnitudes of standardised coefficients so that they sum to unity.

Across all networks, the leading contributors are consistently long-chain EFPs. These are high-degree correlators involving many sequential angular factors, effectively probing the extended angular structure of the jet. Their prominence can be understood from QCD: parton showers are approximately scale-invariant, producing radiation patterns with self-similar angular correlations. An EFP with many edges in a linear chain accumulates information across multiple emissions, making it sensitive to the subtle differences in the radiation pattern of gluons versus quarks. Accordingly, the dominant latent directions are aligned with observables that probe global angular spread and multi-particle correlations.

The stability of these top EFPs across architectures underscores that the models converge on physically meaningful observables. The leading directions in the latent space are not arbitrary vectors in feature space but map onto IRC-safe polynomials with identifiable structures, resonating well with QCD expectations. This provides a bridge between machine-learned representations and theoretical observables, demonstrating that the decision-relevant information encoded in the networks can be traced back to human-understandable physics.

\subsection{Susceptibility to additional soft emission}
\label{subsec:soft_emission_sensitivity}

To probe the sensitivity of the trained networks to additional soft radiation, we construct two controlled variants of each test jet. We add a soft particle with transverse momentum $p_{T}^{\text{soft}} = f \, p_{T}^{\text{Jet}}$ with $f \in \{ 10^{-3}, 10^{-2}, 10^{-1} \}$, where $p_T^{\text{Jet}}$ if the jet's transverse momentum. \emph{Recoiled jets} are produced when the soft particle is placed outside the jet cone and the jet recoils against it. \emph{Soft-added jets} are obtained by inserting the soft particle inside the cone. In the $ f\rightarrow 0$ limit, an IRC-safe classifier should remain insensitive to such perturbations. These controlled modifications test whether the networks are reliant on IRC-unsafe features.

\begin{figure}[t]
	\centering
        \includegraphics[scale=0.5]{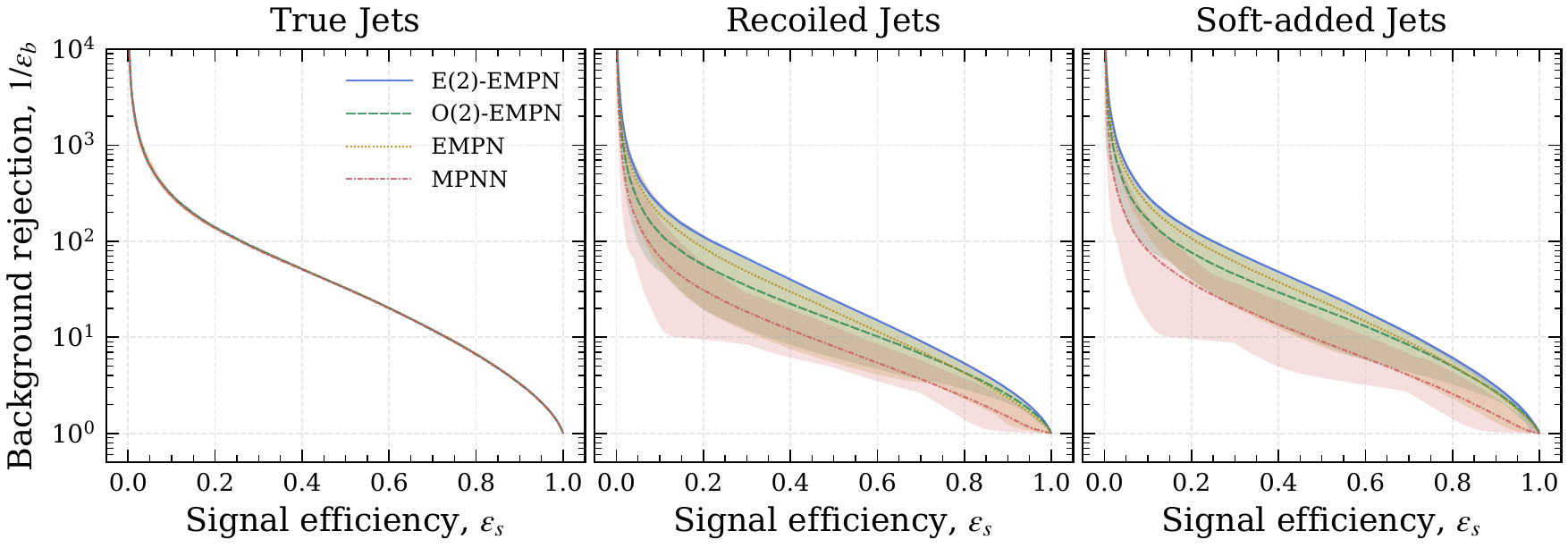}
	\caption{ROC curves for quark vs gluon discrimination, shown as signal efficiency $\epsilon_s$ vs background rejection $1/\epsilon_b$. 
    Left: \emph{true jets}.
    Middle: \emph{recoiled jets} with an additional soft emission outside the jet.
    Right: \emph{soft-added jets} with an additional soft particle inserted inside the jet. For each network, the curves denote the mean over 10 runs, and the shaded bands indicate the minimum-maximum range. In both modified cases, the injected particle has $p_T = 0.01 \, p_{T}^{\text{Jet}}$. }
	\label{fig:roc_pt001}
\end{figure}

At $f=10^{-2}$, the ROC curves in Figure \ref{fig:roc_pt001} show a clear difference between the architectures for both recoiled and soft-added jets. The plot shows the background rejection at fixed signal efficiencies, the curve shows the mean across 10 runs, and the shaded bands indicate the minimum-maximum range. The IRC-safe models retain higher separation than the MPNN, and their run-to-run variation behaves very differently. E(2)-EMPN indicates essentially no visible spread, whereas O(2)-EMPN and EMPN develop modest bands. However, the MPNN has the most degradation in performance with the thickest bands in both cases, indicating greater variability.

\begin{figure}[t]
	\centering
	\includegraphics[scale=0.40]{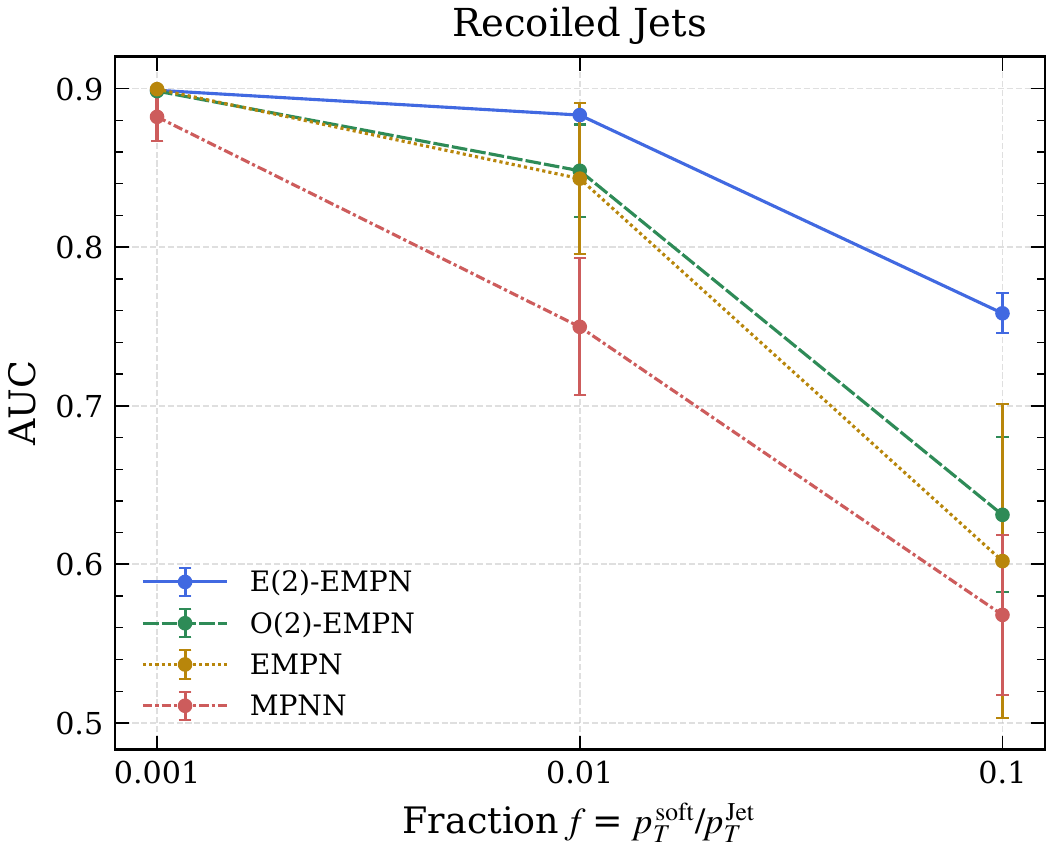}
	\includegraphics[scale=0.40]{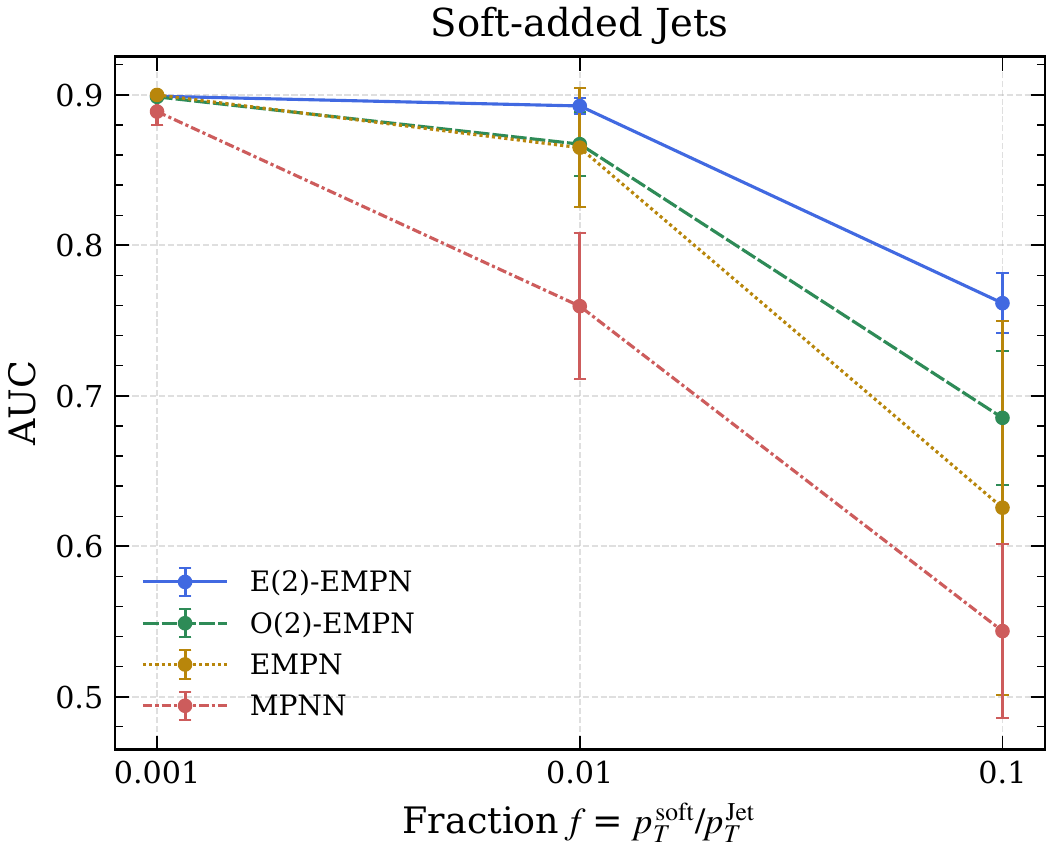}
	\caption{AUC as a function of injected particles transver momentum fraction $f = p_T^{\text{soft}}/p_T^{\text{Jet}}$. Left: Recoiled jets with an additional soft emission outside the jet. Right: Soft-added jet with an additional wide-angle soft particle inside the jet. Markers denote the mean across 10 runs, and error bars indicate one standard deviation.}
	\label{fig:auc_vs_pt}
\end{figure}

Varying the fraction $f$ of soft particles transverse momentum, we plot the AUC values in Figure \ref{fig:auc_vs_pt}. We see that performance for each network declines monotonically with $f$. E(2)-EMPN is consistently the least affected. O(2)-EMPN and EMPN drop more than E(2)-EMPN but remain relatively performant. MPNN has the sharpest decline in performance across both cases. For very soft insertions, with $f=10^{-3}$, the IRC-safe networks are essentially unchanged, while the MPNN already shows a mild but noticeable decrease. The performance separates sharply at $f = 10^{-1}$. The performance order remains E(2)-EMPN, O(2)-EMPN, EMPN, and MPNN in both recoiled and soft-added cases. We observe that the IRC-unsafe MPNN loses almost all separability, with an AUC of nearly 0.5. Among the IRC-safe models, the drop is slightly larger for recoil than for soft particle insertion.

\begin{figure}[t]
	\centering
		\includegraphics[scale=0.5]{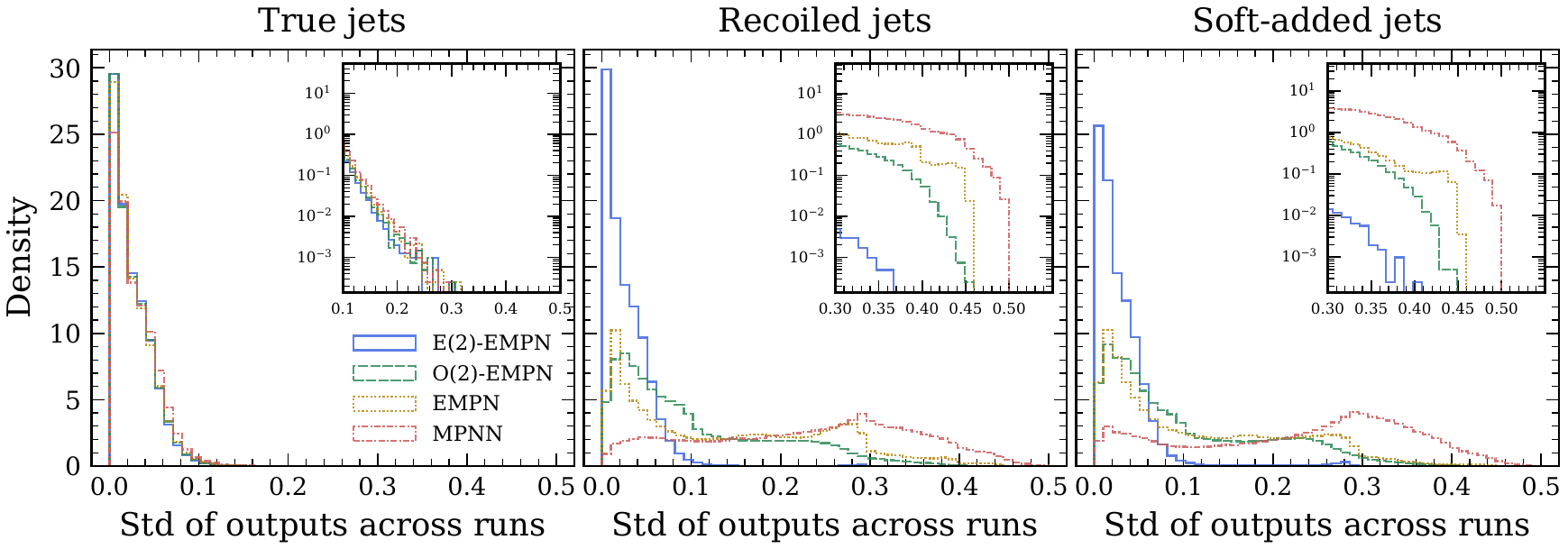}%
	\caption{Histogram of per-jet standard deviation of network output across 10 training instances. Shown separately for \emph{true jets} (left), for \emph{recoiled jets} with emission of additional soft particle outside the jet (middle) and for \emph{soft-added jets} with an additional wide-angle soft particle emitted within the jet (right). Main panels use a linear y-axis, and insets show the tail region on logarithmic scale.}
	\label{fig:output_std}
\end{figure}

To quantify the per-jet stability across runs, Figure \ref{fig:output_std} shows the histogram of the standard deviation of each jet's output across 10 independent runs at $f=10^{-2}$. For a score bounded in $(0,1)$, the theoretical maximum standard deviation is $0.5$, which provides a valuable reference. To display both the bulk behaviour and the tail, the main panels use a linear y-axis, with the inset revealing the tails on a logarithmic scale. On true jets, all models have narrow dispersion with a small high-$\sigma$ tail. However, the spread is still most significant for the MPNN by a minimal amount. Under both perturbations, the tail inflates the most for the MPNN, while E(2)-EMPN remains nearly unchanged, and O(2)-EMPN and EMPN lie in between. These per-jet statistics are consistent with the ROC trends; they show that the E(2)-EMPN is the least sensitive to soft emissions modifications, both in average classification performance and per-sample variability across runs.

A simple geometric picture explains this hierarchy. Recoil effectively shifts the jet's axis, i.e.,  $(y_J, \phi_J)$. The E(2)-EMPN model is invariant under the full Euclidean group in the rapidity-azimuth plane (translations and rotations) and, together with IRC safety, its representation is naturally less sensitive to such global shifts. O(2)-EMPN (only rotational invariance) and EMPN are both IRC-safe but lack explicit translational symmetry and thus show a slightly larger response under axis displacement.\footnote{Note that such recoil sensitivity is dependent on the recombination scheme when the networks take coordinates dependent on the jet axis. For instance, an analysis based on the winner-take-all recombination scheme~\cite{Bertolini:2013iqa,Larkoski:2014uqa}  would generally be less susceptible to recoil effects.} In contrast, the unconstrained MPNN is not IRC-safe and is therefore more susceptible to both types of perturbations. On the unmodified test jets, all four models perform similarly. However, the apparent differences emerge when we apply these modifications, where they have varying effects once the jets are perturbed in controlled ways. 

Taken together, these tests separate architectures that learn physics-aligned decision surfaces from those that achieve similar baseline AUCs but lean on more fragile cues. Models with appropriate symmetry priors preserve their behaviour under perturbations that QCD deems uninformative.

\section{Summary and Conclusions}
\label{sec:conclusions}

We investigated the interpretability of graph neural networks applied to the task of quark-gluon discrimination, with a particular focus on the role of physics-inspired inductive biases. We designed message passing architectures that explicitly incorporate two key constraints: IRC safety and equivariance under transformations in the rapidity-azimuth plane, specifically E(2) and O(2) symmetry. These choices ensure that the networks learn features consistent with fundamental properties of QCD and with the symmetries of collider detectors. For comparison, we also trained baseline networks without these constraints, allowing us to assess the impact of the inductive biases in a controlled way.

We performed a detailed evaluation of the classification performance of these models on a large dataset of simulated jets. All networks achieved competitive discriminating power, with AUC values close to 0.9. However, significant differences emerged in their stability and robustness. When subjected to controlled modifications of the input, such as the addition of soft particles or recoils, the IRC-safe networks maintained stable predictions. In contrast, the IRC-unsafe baseline was more strongly affected. These observations confirm that embedding physical constraints not only matches but can surpass unconstrained networks in terms of reliability and resilience.

Beyond classification performance, we examined how the different architectures organise information internally. Using principal component analysis of the graph-level latent representations, we found a striking structural difference between equivariant and non-equivariant models. In the non-equivariant models, the explained variance increases almost identically regardless of IRC safety. Similarly, regardless of the underlying group, both equivariant and IRC-safe models had a similar and comparatively faster growth of explained variance as a consequence of the continuous symmetry constraints in the latent space. 

To further connect the learned representations to known physics, we regressed Energy Flow Polynomials, which form an overcomplete basis of IRC-safe jet observables, onto the leading principal components for each model. For the networks constrained by IRC safety and equivariance, the regression achieved high coefficients of determination, demonstrating that their dominant latent features can be directly mapped onto established jet observables. The unconstrained model, by contrast, showed weaker correspondence, indicating that its dominant features are less well aligned with known, interpretable quantities. This analysis provides concrete evidence that the inclusion of inductive biases leads networks to learn representations that are not only robust but also more physically transparent.

Taken together, this study demonstrates that symmetry- and safety-aware network design provides a principled approach to enhancing the interpretability of deep learning methods in collider physics. By aligning the latent spaces of GNNs with established QCD observables and demonstrating the structural advantages of physics-constrained representations, we have provided a framework for bridging the gap between high-performing machine learning models and the analytic tools traditionally used in jet physics. Our findings indicate that respecting physical principles in model design is not merely a theoretical preference but a practical route toward stable, reliable, and interpretable machine learning applications at colliders.

\section*{Acknowledgements}
Research work at the Physical Research Laboratory (PRL) is funded by the Department of
Space, Government of India. The authors gratefully acknowledge the resources provided by the
Param Vikram-1000 High Performance Computing Cluster and the TDP project infrastructure at
PRL, which were extensively used for computational aspects of this work. STFC supports VSN and MS under grant ST/X003167/1.

\appendix 
\section{Group action, orbits, and equivariance}
\label{app:equiv}
Let $\mathcal{G}$ be a group with identity element $e$. A (left) action of $\mathcal{G}$ on a set $X$ is a map $ \rho_X : \mathcal{G} \times X \to X$, such that for all $g_1, g_2 \in \mathcal{G}$ and $x \in X$ we have,
\begin{align}
	\rho_X(e,x) &= x \\
	\rho_X \bigl(g_1,\,\rho_X(g_2,x)\bigr)&=\rho_X(g_1g_2,\,x).
\end{align}

For $x\in X$, the \emph{orbit} of $x$ is the set of points reachable from $x$ by the action of $\mathcal{G}$.
\begin{equation}
	\operatorname{Orb}_\mathcal{G}(x) \coloneqq \{\, \rho_X(g, x) : g\in \mathcal{G} \,\}.
\end{equation}
Orbits collect points that are the same up to the group action.  

%\textbf{Orbits as equivalence classes.}
Declare $x \sim x'$ if $ \exists \, g \in \mathcal{G}$ with $x' = \rho_X (g, x)$. This defines an equivalence relation $\sim$ on $X$, and we get equivalence classes, which are the orbits.
\begin{equation}
	[x]_\sim \;=\; \operatorname{Orb}_\mathcal{G}(x)\quad.
\end{equation}
Consequently, the collection of all orbits forms a \emph{partition} of $X$. Hence, the action partitions $X$ into disjoint sets, and every point lies in exactly one orbit.

The group action is \emph{transitive} if $X$ consists of a single orbit, i.e.\ for all $x, x' \in X$ there exists $g \in \mathcal{G}$ such that $x' = \rho_X(g, x)$. In other words, any point can be moved to any other by some group element.

Given a function $f:X \to Y$, the \emph{fibre} of an element $y \in \im(f)$ is
\begin{equation}
	f^{-1}(y) \coloneqq \{ \, x \in X : f(x)=y \, \}\quad.
\end{equation}
Fibres over the entire image of $f$ also partition $X$ and form equivalence classes.

In the context of $\mathcal{G}$-action on $X$, there is a canonical projection map that maps orbits to their equivalence class
\begin{equation}
	\pi_X:X \to X/\mathcal{G},\qquad \pi_X(x)\coloneqq \operatorname{Orb}_{\mathcal{G}}(x),
\end{equation}
where $\pi_X$ is the \emph{orbit map}, and $X/\mathcal{G}$ denotes the set of orbits. Thus, the partition of $X$ into orbits is the fibre decomposition of $\pi_X$.

Given the actions of $\mathcal{G}$ on the domain $X$ and the codomain $Y$, a function $f: X \to Y$ is \emph{$\mathcal{G}$-equivariant} if it commutes with the group action, i.e.,

\begin{equation}
	f \bigl( \rho_X(g,x) \bigr) = \rho_Y \bigl(g,\,f(x) \bigr) \quad \forall g\in \mathcal{G}, \; x\in X.
\end{equation}

If the action on $Y$ is trivial (i.e.\ $\rho_Y(g,y)=y$ for all $g\in \mathcal{G}$ and $y\in Y$), then $f$ is \emph{$\mathcal{G}$-invariant}. Equivalently,
\begin{equation}
	f \bigl( \rho_X(g,x) \bigr) = f(x) \quad \forall \, g,x.
\end{equation}
Under equivariance, points in the same orbit of $X$ map to a single orbit of $Y$. Under invariance, the orbit of an element of $Y$ is the point itself, which results in the function being constant on the orbits in $X$.

\section{Principal component analysis}
\label{app:pca} 
Principal Component Analysis (PCA) is a lossy, linear dimensionality reduction method. It re-expresses the data in a new orthonormal basis, whose axes are ordered by the variance they explain. 

Let $X \in \mathbb{R}^{m \times n}$ be the data matrix, with $m$ samples of $n$ features each. The columns of the data are first mean-centred, giving $\tilde X$. The covariance matrix is 
\begin{equation}
	\Sigma = \frac{1}{m-1}\,\tilde X^{\top}\tilde X \in \mathbb{R}^{n \times n}.
\end{equation}

An eigendecomposition of this covariance matrix is performed to obtain a diagonal matrix $S$. 
\begin{equation}
	\Sigma = V S V^{\top}, \qquad S = \operatorname{diag}(\lambda_{1},\lambda_{2},\ldots,\lambda_{n}),
\end{equation}
where the eigenvalues satisfy $\lambda_{1} \ge \lambda_{2} \ge \cdots \ge \lambda_{n}$. The columns of $V$ are the principal directions. The variance explained by the $i$-th component is $\lambda_i$, and its proportion is $\frac{\lambda_i}{\sum_{j=1}^{n}\lambda_j}$.

To reduce the data to $k$ dimensions ($k \le n$), keep the top $k$ directions $V_k \in \mathbb{R}^{n \times k}$, then the reduced features are $Z = \tilde X V_k \in \mathbb{R}^{m \times k}$.

In the text, the term \emph{principal components} is used to describe the projection of the data onto these axes. The working assumption behind PCA is that directions with higher variance tend to encode the relevant structure. This yields a compact representation that preserves the dominant patterns of the data.

\bibliographystyle{JHEP}
\bibliography{ref}

\end{document}